\documentclass[runningheads]{llncs}
\usepackage[T1]{fontenc}
\usepackage{graphicx,verbatim}
\usepackage{booktabs}
\usepackage{multirow}
\usepackage{amssymb}
\usepackage{algorithm}
\usepackage{algpseudocode}
\usepackage{enumitem}
\usepackage{hyperref}

\begin{document}
\title{CardiacFlow: 3D+t Four-Chamber Cardiac Shape Completion and Generation via Flow Matching}
\titlerunning{CardiacFlow}

\author{Qiang Ma\inst{1} \and
Qingjie Meng\inst{2,3} \and
Mengyun Qiao\inst{1,4} \and
Paul M. Matthews\inst{1,5,6} \and\\
Declan P. O'Regan\inst{7} \and
Wenjia Bai\inst{1,3,4}}

\authorrunning{Q. Ma et al.}
% First names are abbreviated in the running head.
% If there are more than two authors, 'et al.' is used.
%
\institute{Department of Brain Sciences, Imperial College London, London, UK \and
School of Computer Science, University of Birmingham, Birmingham, UK \and
Department of Computing, Imperial College London, London, UK \and
Data Science Institute, Imperial College London, London, UK \and
UK Dementia Research Institute, Imperial College London, London, UK \and
Rosalind Franklin Institute, Harwell Science and Innovation Campus, Didcot, UK \and
MRC Laboratory of Medical Sciences, Imperial College London, London, UK\\
\email{q.ma20@imperial.ac.uk}}

\maketitle              % typeset the header of the contribution
\begin{abstract}
Learning 3D+t shape completion and generation from multi-view cardiac magnetic resonance (CMR) images requires a large amount of high-resolution 3D whole-heart segmentations (WHS) to capture shape priors. In this work, we leverage flow matching techniques to learn deep generative flows for augmentation, completion, and generation of 3D+t shapes of four cardiac chambers represented implicitly by segmentations. Firstly, we introduce a latent rectified flow to generate 3D cardiac shapes for data augmentation, learnt from a limited number of 3D WHS data. Then, a label completion network is trained on both real and synthetic data to reconstruct 3D+t shapes from sparse multi-view CMR segmentations. Lastly, we propose CardiacFlow, a novel one-step generative flow model for efficient 3D+t four-chamber cardiac shape generation, conditioned on the periodic Gaussian kernel encoding of time frames. The experiments on the WHS datasets demonstrate that flow-based data augmentation reduces geometric errors by 16\% in 3D shape completion. The evaluation on the UK Biobank dataset validates that CardiacFlow achieves superior generation quality and periodic consistency compared to existing baselines. The code of CardiacFlow is released publicly at \url{https://github.com/m-qiang/CardiacFlow}.

% Authors must provide keywords and are not allowed to remove this Keyword section.

\end{abstract}
\section{Introduction}

As the current gold standard imaging modality for assessing cardiac structure and function, cine cardiac magnetic resonance (CMR) imaging facilitates the quantitative evaluation of cardiac phenotypes, such as chamber volumes and ejection fractions \cite{Petersen2016UK}. The reconstruction of 3D+t cardiac shape models \cite{Deng2023,Fonseca2011,Galazis2025,meng2023deepmesh,Xia2022} from CMR images plays an essential role in understanding regional patterns of cardiac diseases \cite{Biffi2018} and conducting electrophysiological and biomechanical studies of the heart \cite{Niederer2019,Roney2019,Sermesant2012,Strocchi2020}. However, it is a challenging inverse problem to reconstruct high-resolution four-chamber shapes of the heart from sparse multi-view CMR sequences. The four cardiac chambers, including the left ventricle (LV), right ventricle (RV), left atrium (LA), and right atrium (RA), are delineated by stacks of 2D image slices acquired from multiple view planes. Although there are multiple slices covering the ventricles, the atria are only covered by one or two slices. The challenges also come from potential motion artefacts and slice misalignment caused by respiratory motion during image acquisition.

Previous works demonstrated the feasibility of leveraging shape priors from 3D whole-heart segmentations (WHS) of CT scans \cite{scot2018coronary} to learn 3D+t cardiac shape completion in CMR segmentations \cite{Muffoletto2024,xu2023whole,Xu2023,Xu2024}. However, this requires a large amount of 3D WHS data, which are not always available due to data regulation issues and the expensive cost of image acquisition and annotation. Existing public WHS datasets \cite{kiricsli2013casdc,metz2009ccec,tobon2015lasc,Zhuang2019} only contain dozens of 3D segmentations, which highlight the necessity of data augmentation. Deep generative models \cite{heusel2017fid,karras2019stylegan,kingma2013vae} offer a potential solution to enrich the quantity and variety of 3D cardiac shapes. While deep generative models have been utilised to characterise 3D shape distribution \cite{Kong2024} or spatio-temporal motion patterns of the heart \cite{Qiao2023,sorensen2024stndf}, their applications to augmenting, completing and generating 3D+t four-chamber cardiac shapes are still less explored.

Recently, diffusion models have achieved great success in data generation \cite{ho2020ddpm,song2020score}. For efficient training and sampling, latent diffusion model \cite{rombach2022ldm} operates in the latent space and thus is naturally suitable for 3D medical images. Compared to diffusion models that require considerable training data and thousands of stochastic sampling steps \cite{ho2020ddpm,rombach2022ldm,wang2023patch}, latest flow matching approaches \cite{esser2024stable3,lipman2022fm,liu2022rectified} are data- and time-efficient by learning a continuous normalising flow between two distributions. In particular, a rectified flow technique \cite{esser2024stable3,lee2025rectified,liu2022rectified} is developed to learn a straight optimal transport path from noise to data, enabling it to capture the target distribution and generate data samples in a few steps.

In this work, we propose a novel solution to learn 3D+t four-chamber cardiac shape completion and generation via flow matching, in which the shapes are represented implicitly by segmentation maps. The main contributions of this work are listed as follows:
\begin{itemize}[leftmargin=*]
\item We introduce a latent rectified flow to generate synthetic 3D shapes of four cardiac chambers, learned from a limited number of 3D WHS data via flow matching. Such data augmentation significantly improves the performance of 3D shape completion.
\item A label completion network is trained on both real and synthetic 3D shapes, and further applied to reconstruct accurate 3D+t cardiac shapes from multi-view CMR segmentations on the UK Biobank dataset for a large population.
\item We propose \textit{CardiacFlow}, an efficient one-step generative flow model for 3D+t four-chamber cardiac shape generation. Incorporating periodic Gaussian kernel encoding of time frames and learnable initial values, CardiacFlow achieves state-of-the-art generation quality and periodic consistency.
\end{itemize}

\section{Methods}
\subsection{Latent Rectified Flow for Data Augmentation} \label{sec:2-1}
Learning to complete four-chamber cardiac shapes from multi-view segmentations requires a large amount of training data to capture the shape variation. Given a limited number of 3D WHS data \cite{kiricsli2013casdc,metz2009ccec,tobon2015lasc,Zhuang2019}, we introduce \textit{flow matching} (FM) \cite{lipman2022fm} to generate 3D cardiac shapes for data augmentation. \newline

\noindent\textbf{Flow Matching.} Given a noise distribution $p(x_0)=\mathcal{N}(x_0;0,I)$ and data distribution $q(x_1)$, FM models a generative process by learning a \textit{continuous normalising flow} from noise $x_0\sim p(x_0)$ to data $x_1\sim q(x_1)$ defined by an ODE:
\begin{equation}\label{eq:cnf}
dx_t=v_t(x_t;\theta)dt,~x_0\sim p(x_0),~t\in[0,1],
\end{equation}
where $v_t$ is a \textit{time-varying vector field} parameterised by a neural network with learnable parameters $\theta$. It defines a probability path $p_t$ with $x_t\sim p_t(x)$. Ideally, the path $p_t$ should flow from noise to data such that $p_0(x)=p(x_0)$ and $p_1(x)=q(x_1)$. 
To learn the flow (\ref{eq:cnf}), a conditional FM loss function is defined as:
\begin{equation}\label{eq:cfm-loss}
\mathcal{L}_{\mathrm{CFM}}(\theta)=\mathbb{E}_{t,q(x_1),p_t(x|x_1)}\left[\|v_t(x_t;\theta)-u_t(x_t;x_1)\|^2\right].
\end{equation}
Such a loss trains the flow (\ref{eq:cnf}) to match a conditional flow defined by a vector field $u_t$ conditioned on data $x_1$, where $u_t$ is usually constructed as an optimal transport path $u_t(x_t;x_1)=(x_1-x_t)/(1-t)$, \emph{i.e.}, the shortest straight path from noise $x_0$ to data $x_1$ \cite{lipman2022fm,liu2022rectified}. In this case, the flow (\ref{eq:cnf}) is also called a \textit{rectified flow}. After training, new data samples can be generated by integrating the ODE (\ref{eq:cnf}).
\newline

\begin{figure}[t]
\centering
\includegraphics[width=1.0\linewidth]{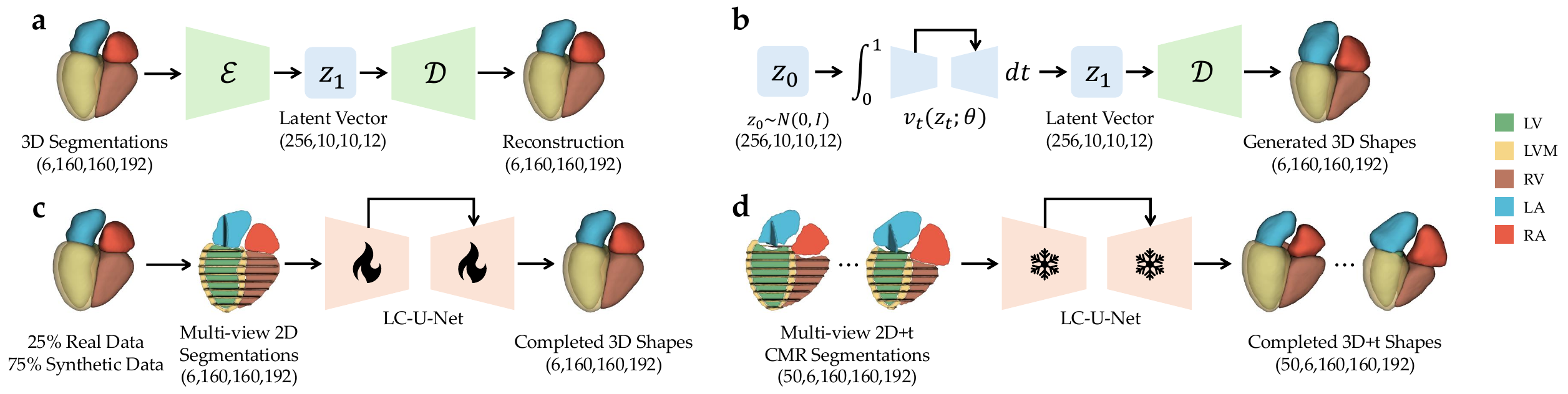}
\caption{The procedure of 3D+t cardiac shape completion. (a) An autoencoder is trained to extract latent vectors from 3D shapes. (b) A LRF is learned to generate 3D shapes for data augmentation. (c) A LC-U-Net is trained on both real and synthetic data for 3D shape completion. (d) The trained LC-U-Net is applied to 3D+t shape completion.}
\label{fig:shape_completion}
\end{figure}

\noindent\textbf{Latent Rectified Flow.} Instead of learning high-dimensional distributions, FM can be performed efficiently in a latent space \cite{esser2024stable3}. As shown in Fig. \ref{fig:shape_completion}-a,b, we train an autoencoder $\hat{x}=\mathcal{D}(\mathcal{E}(x))$ with encoder $\mathcal{E}$ and decoder $\mathcal{D}$ to extract latent vectors $z_1=\mathcal{E}(x)$ from 3D cardiac segmentations $x\sim q(x)$. Then, a \textit{latent rectified flow} (LRF) is defined as $dz_t=v_t(z_t;\theta)dt$, where the initial value $z_0\sim\mathcal{N}(0,I)$ has the same dimensionality as the latent vector $z_1$, and $v_t$ is parameterised by a 3D flow U-Net \cite{ho2020ddpm,rombach2022ldm,ronneberger2015unet} with \textit{adaptive instance normalisation} layers \cite{karras2019stylegan}. 

To train the LRF, we randomly sample $t\in[0,1]$, $z_0\sim\mathcal{N}(0,I)$, and $z_1=\mathcal{E}(x)$ with $x\sim q(x)$ for each iteration. In order to match the LRF to a straight optimal transport path, $z_t$ is approximated by a linear interpolation $z_t=(1-t)z_0+tz_1$ between $z_0$ and $z_1$. Hence, the latent FM loss is reformulated as \cite{esser2024stable3,liu2022rectified}:
\begin{equation}\label{eq:lrf-loss}
\mathcal{L}_{\mathrm{LFM}}(\theta)=\mathbb{E}_{t,q(x),p_t(z|z_1)}\left[\|v_t((1-t)z_0+tz_1;\theta)-(z_1-z_0)\|^2\right].
\end{equation}
After training, we sample $z_0$ and integrate the LRF by forward Euler method with $T$=100 steps, in which the flow U-Net $v_t$ is evaluated repeatedly, to generate a latent vector $\hat{z}_1$. The 3D segmentation is reconstructed by $\hat{x}_1=\mathcal{D}(\hat{z}_1)$ as shown in Fig. \ref{fig:shape_completion}-b. Note that if the flow trajectory is ideally straight, the forward Euler method can provide exact ODE solution within one step \cite{liu2022rectified}. In spite of learning from a limited number of WHS data \cite{kiricsli2013casdc,metz2009ccec,tobon2015lasc,Zhuang2019}, the LRF is data efficient and able to generate diverse 3D cardiac shapes for effective data augmentation.

\subsection{3D+t Four-Chamber Cardiac Shape Completion}\label{sec:2-2}

The data augmentation performed by LRF enables us to capture accurate shape distribution for 3D+t shape completion. Given augmented 3D cardiac segmentations, a label completion U-Net (LC-U-Net) \cite{ronneberger2015unet,xu2023whole,Xu2023} is trained to reconstruct 3D cardiac segmentations from multi-view 2D segmentations (Fig. \ref{fig:shape_completion}-c). The input multi-view 2D segmentations, including short-axis (SAX) and long-axis (LAX) views, are synthesised
following \cite{xu2023whole,Xu2023,Xu2024}, incorporating in-plane motion between SAX slices and misalignment between SAX and LAX views. The displacements of the slices are simulated as a Gaussian distribution $\mathcal{N}(0,\lambda^2 I)$, where $\lambda$ is the corruption level following a uniform distribution between 0 and 4 mm. For each training epoch, the training data consist of 25\% real data collected from the WHS datasets \cite{kiricsli2013casdc,metz2009ccec,tobon2015lasc,Zhuang2019} and 75\% synthetic 3D shapes generated by the LRF. After training, as illustrated in Fig. \ref{fig:shape_completion}-d, we apply the LC-U-Net to complete four-chamber cardiac shapes across $M$=50 time frames of multi-view CMR segmentations on the UK Biobank (UKB) database \cite{Petersen2016UK}, curating a 3D+t four-chamber cardiac shape dataset for a large population.

\subsection{CardiacFlow for 3D+t Cardiac Shape Generation}\label{sec:2-3}

\begin{figure}[t]
\centering
\includegraphics[width=1.0\linewidth]{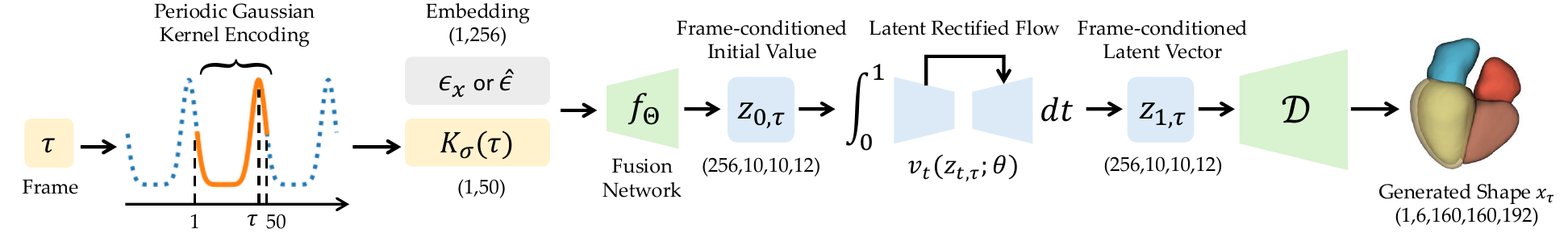}
\caption{The architecture of CardiacFlow. Each frame $\tau$ is encoded by a PGK $\mathcal{K}_{\sigma}(\tau)$ and fused with a learnable embedding $\epsilon_x$ (train) or a sampled embedding $\hat{\epsilon}$ (test). The fusion network predicts a frame-conditioned initial value $z_{0,\tau}$ for the LRF to generate a latent vector $z_{1,\tau}$, which is decoded to a 3D shape $x_{\tau}=\mathcal{D}(z_{1,\tau})$.}
\label{fig:cardiacflow}
\end{figure}

Based on the curated 3D+t cardiac shape dataset, we develop \textit{CardiacFlow}, an efficient one-step generative flow that extends LRF to generate spatio-temporal shapes of four cardiac chambers. Given completed 3D+t segmentations $x\sim q(x)$ from the UKB dataset, we train a 3D autoencoder $\hat{x}_{\tau}=\mathcal{D}(\mathcal{E}(x_{\tau}))$ to learn a latent vector $z_{1,\tau}=\mathcal{E}(x_{\tau})$ from 3D shape $x_{\tau}$ for time frame $\tau\in\{1,...,M\}$. As shown in Fig. \ref{fig:cardiacflow}, CardiacFlow defines a LRF from a learnable frame-conditioned initial value $z_{0,\tau}$ to the latent vector $z_{1,\tau}$ for each frame $\tau$, modelled by an ODE:
\begin{equation}\label{eq:cardiacflow}
dz_{t,\tau}=v_t\left(z_{t,\tau};\theta\right)dt,~z_{0,\tau}=f_{\Theta}\left(\epsilon_x,\mathcal{K}_{\sigma}(\tau)\right),~t\in[0,1],
\end{equation}
where $v_{t}$ is parameterised by a 3D U-Net in latent space, $f_{\Theta}$ is a fusion network, $\epsilon_x$ is a learnable embedding for each training data, and $\mathcal{K}_\sigma(\tau)$ is a \textit{periodic Gaussian kernel} (PGK) \textit{encoding} of time frame $\tau$.
\newline

\noindent\textbf{Periodic Gaussian Kernel Encoding.} For cardiac 3D+t shape generation, the heart should have a consistent shape at the start ($\tau$=1) and end ($\tau$=$M$) of a cardiac cycle. However, existing generative models \cite{Qiao2023,sorensen2024stndf} cannot preserve such periodic consistency. Instead of using scalar or one-hot conditioning variable, CardiacFlow employs a PGK $\mathcal{K}_{\sigma}:\mathbb{R}\rightarrow\mathbb{R}^M$ to encode time frame $\tau$, based on the periodicity of heartbeat. The $m$-th element of the vector $\mathcal{K}_{\sigma}(\tau)$ is defined by a Gaussian kernel $[\mathcal{K}_{\sigma}(\tau)]_m\triangleq\frac{1}{\sqrt{2\pi}\sigma}\exp(-\frac{d(m,\tau)^2}{2\sigma^2})$, where $d(m,\tau)\triangleq|\mathrm{mod}(m-\tau+M/2, M)-M/2|$ is a distance metric, which is symmetric $d(m, \tau)=d(\tau, m)$ and periodic $d(m+M,\tau)=d(m,\tau)$ with period $M$. As depicted in Fig. \ref{fig:cardiacflow}, the elements of $\mathcal{K}_{\sigma}(\tau)$ follow a truncated and periodic Gaussian distribution, where $[\mathcal{K}_{\sigma}(\tau)]_m$ achieves maximum when $m=\tau$. For each frame $\tau$, the PGK encoding allows CardiacFlow to capture the information of neighbouring frames within period $M$ and guarantees the periodic consistency in a cardiac cycle.
\newline

\noindent\textbf{Learnable Initial Value.} CardiacFlow aims to generate each frame of the 3D+t cardiac shapes in one step. One key factor is to rectify and straighten the flow path, such that one step of numerical integration can provide accurate ODE solution \cite{lipman2022fm,liu2022rectified}. While FM samples random coupling of noise and data for training \cite{lipman2022fm,liu2022rectified}, inspired by \cite{Kong2024,sorensen2024stndf}, we assign a low-dimensional learnable embedding $\epsilon_x$ to each training data $x\sim q(x)$. A fusion network $f_{\Theta}$ with multiple MLP layers is trained, such that the learnable embedding $\epsilon_x$ is fused with the frame encoding $\mathcal{K}_{\sigma}(\tau)$ and upsampled to the same size as the latent vector $z_{1,\tau}$. It provides a learnable frame-conditioned initial value $z_{0,\tau}=f_{\Theta}\left(\epsilon_x,\mathcal{K}_{\sigma}(\tau)\right)$ for the ODE (\ref{eq:cardiacflow}), and therefore CardiacFlow can optimise the initial value and flow trajectory simultaneously to learn a straight optimal transport path for one-step generation. For data generation, the learnable embedding $\epsilon_x$ is replaced by $\hat{\epsilon}\sim\mathcal{N}(\mu_{\epsilon}, \Sigma_{\epsilon})$, an empirical Gaussian distribution defined by $\{\epsilon_x\}_{x\sim q(x)}$.
\newline

\noindent\textbf{Beta Sampling.} 
Another factor for one-step generation is the sampling strategy \cite{esser2024stable3,lee2024beta,lee2025rectified}. FM typically samples the time steps $t\in[0,1]$ uniformly during training \cite{lipman2022fm,liu2022rectified}, providing very few samples at $t$=0 for one-step generation from initial values \cite{esser2024stable3,lee2024beta,lee2025rectified}. To alleviate this issue, CardiacFlow samples time $t$ from a \textit{Beta distribution}, $\mathrm{Beta}(a,b)$, defined on $[0,1]$ with $a$=0.1 and $b$=2.0 \cite{lee2024beta}. The density function monotonically decreases so that more samples are drawn near $t$=0.

The training and generation procedures of CardiacFlow are summarised in Algorithms \ref{alg:training} and \ref{alg:generation}. The flow ODE in Algorithm \ref{alg:generation} is solved by one step of forward Euler, \emph{i.e.}, $\hat{z}_{1,\tau}=\hat{z}_{0,\tau}+v_{0}(\hat{z}_{0,\tau};\theta)$. Note that the Beta sampling used in training also allows few-step generation, which enhances the quality of generated shapes by more integration steps. The 3D shape $\hat{x}_{\tau}=\mathcal{D}(\hat{z}_{1,\tau})$ at each frame $\tau$ is reconstructed by decoding the generated latent vector. The 3D+t cardiac shapes $\hat{x}$ are generated by iterating over all time frames $\tau=1,...,M$.

\algrenewcommand\algorithmicindent{1em}%
\begin{figure}[t]
\begin{minipage}[t]{0.57\textwidth}
\begin{algorithm}[H]
% \scriptsize
\caption{Training}\label{alg:training}
\begin{algorithmic}[1]
\State \textbf{input:} 3D+t shapes $q(x)$, embeddings $\epsilon_x$
% \Repeat
\State \textbf{repeat}
\State \hspace{\algorithmicindent}
$x\sim q(x),~\tau\sim\mathrm{Uniform}(\{1,...,M\})$
\State \hspace{\algorithmicindent}
$z_{1,\tau}=\mathcal{E}(x_{\tau}),~z_{0,\tau}=f_{\Theta}\left(\epsilon_x,\mathcal{K}_{\sigma}(\tau)\right)$
\State \hspace{\algorithmicindent}
$t\sim\mathrm{Beta}(a,b),~z_{t,\tau}=(1-t)z_{0,\tau}+tz_{1,\tau}$
\State \hspace{\algorithmicindent}
$\mathcal{L}(\theta,\Theta,\epsilon_x)=\|v_{t}(z_{t,\tau};\theta)-(z_{1,\tau}-z_{0,\tau})\|^2$
\State \hspace{\algorithmicindent}
update $\theta$, $\Theta$, $\epsilon_x$
\end{algorithmic}
\end{algorithm}
\end{minipage}
\hfill
\begin{minipage}[t]{0.42\textwidth}
\begin{algorithm}[H]
\caption{Generation} \label{alg:generation}
\begin{algorithmic}[1]
% \small
% \State \textbf{input:} integration steps $T$, step size $h$
\State $\hat{\epsilon}\sim\mathcal{N}(\mu_{\epsilon},\Sigma_{\epsilon})$
\For{$\tau=1,...,M$}
\State $\hat{z}_{0,\tau}=f_{\Theta}\left(\hat{\epsilon},\mathcal{K}_{\sigma}(\tau)\right)$
\State $\hat{z}_{1,\tau}=\hat{z}_{0,\tau}+\int_{0}^{1}v_{t}(\hat{z}_{t,\tau};\theta)dt$
\State $\hat{x}_{\tau} = \mathcal{D}(\hat{z}_{1,\tau})$
\EndFor
\State \textbf{return} $\hat{x}=\{\hat{x}_{\tau}\}$ %_{\tau=1,...,M}$
\end{algorithmic}
\end{algorithm}
\end{minipage}
\end{figure}

\begin{table}[t]\scriptsize
\centering
\setlength\tabcolsep{2.0pt}
\caption{The results of 3D shape completion. A LC-U-Net trained without synthetic data is compared to LC-U-Nets with data augmentation using VAE or LRF. 
The bolded results are significantly better than other methods (t-test, $p<0.05$).}
\label{tab:shape_completion}
\begin{tabular}{clccccc}
\toprule
 & Method & LV & LVM & RV & LA & RA \\
\midrule
\multirow{3}{*}{\begin{tabular}{@{}c@{}}HD95$\downarrow$\\(mm)\end{tabular}}
& LC-U-Net  &  2.084$\pm$0.630  &  
1.995$\pm$0.561  &  2.435$\pm$1.007  &  
3.120$\pm$1.010  &  4.385$\pm$1.919 \\
& LC-U-Net (VAE) &  1.882$\pm$0.606  &  
1.870$\pm$0.601  &  2.155$\pm$1.161  &  
2.948$\pm$0.986  &  3.999$\pm$1.987 \\
& LC-U-Net (LRF) &  
\textbf{1.769}$\pm$\textbf{0.521}  &  
\textbf{1.693}$\pm$\textbf{0.492}  &  
\textbf{2.024}$\pm$\textbf{0.982}  &  
\textbf{2.675}$\pm$\textbf{0.884}  &  
\textbf{3.581}$\pm$\textbf{1.697} \\
\midrule
\multirow{3}{*}{DSC$\uparrow$}
& LC-U-Net  &  0.955$\pm$0.017  & 
0.900$\pm$0.027  &  0.944$\pm$0.023  &  
0.929$\pm$0.021  &  0.905$\pm$0.039 \\
& LC-U-Net (VAE) &  0.960$\pm$0.014  &  
0.909$\pm$0.024  &  0.952$\pm$0.022  &  
0.934$\pm$0.021  &  0.911$\pm$0.041 \\
& LC-U-Net (LRF) &
\textbf{0.963}$\pm$\textbf{0.012}  &  
\textbf{0.917}$\pm$\textbf{0.021}  &
\textbf{0.954}$\pm$\textbf{0.020}  &  
\textbf{0.939}$\pm$\textbf{0.018}  &
\textbf{0.918}$\pm$\textbf{0.036} \\
\bottomrule
\end{tabular}
\end{table}

\section{Experiments}
\noindent\textbf{Dataset.} For 3D shape augmentation and completion, we collect 160 WHS data from multiple public datasets including WHS++ \cite{Zhuang2019} and other challenges \cite{kiricsli2013casdc,metz2009ccec,tobon2015lasc}. All segmentation maps are rigidly aligned to a WHS atlas \cite{Zhuang2016} and clipped to the size of 160$\times$160$\times$192. The dataset is randomly split by the ratio of 6/1/3 for training, validation and test. The 3D+t shape completion and generation are performed on 1,000 UKB CMR scans  \cite{Petersen2016UK}, which are split by the ratio of 6/1/3 as well. The 2D+t segmentations in SAX, two- and four-chamber LAX views are generated using a publicly available method \cite{bai2018automated} with manual quality control, including the classes of LV, LV myocardium (LVM), RV, LA and RA. All 2D slices are mapped into 3D space and rigidly aligned to the WHS atlas \cite{Zhuang2016}. All experiments are conducted on a Nvidia 3080 GPU with 12GB memory.\newline

\noindent\textbf{3D Shape Completion.} We examine the effects of data augmentation via LRF on 3D cardiac shape completion on the WHS datasets \cite{kiricsli2013casdc,metz2009ccec,tobon2015lasc,Zhuang2019}. We train a LRF for 1,000 epochs and it takes 0.266s to generate a 3D segmentation with $T$=100 steps. A VAE \cite{kingma2013vae} is trained for comparison. We also train a latent diffusion model \cite{rombach2022ldm}, which however, fails to converge with limited number of training data. Then, we train LC-U-Nets for 600 epochs with data augmentation, which includes 25\% real data and 75\% synthetic data generated online by VAE \cite{kingma2013vae} or LRF for each epoch, compared to a LC-U-Net trained without synthetic data. For each of 48 3D segmentations in the test set, we simulate 10 different multi-view 2D segmentations at random motion level $\lambda$, resulting in 480 test samples for evaluation. The Dice coefficient (DSC) and 95th percentile of Hausdorff distance (HD95) are measured and reported in Table \ref{tab:shape_completion}. The statistical significance is examined by paired t-test. Compared to the LC-U-Net trained without data augmentation, the LRF enhances the data diversity and reduces the geometric errors significantly by 16.2\% in terms of HD95.

\begin{figure}[t]\scriptsize
\centering
\begin{minipage}[t]{.49\linewidth}
\centering
\includegraphics[width=1.0\linewidth]{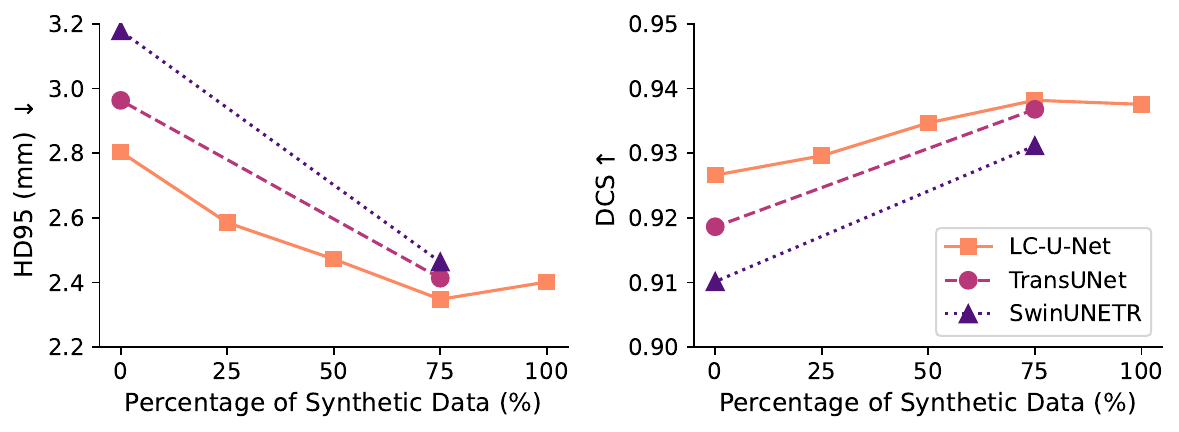}
\caption{Shape completion with different ratio of synthetic data and neural networks.}
\label{fig:data_augment}
\end{minipage}
\hfill
\begin{minipage}[t]{.49\linewidth}
\centering
\includegraphics[width=1.0\linewidth]{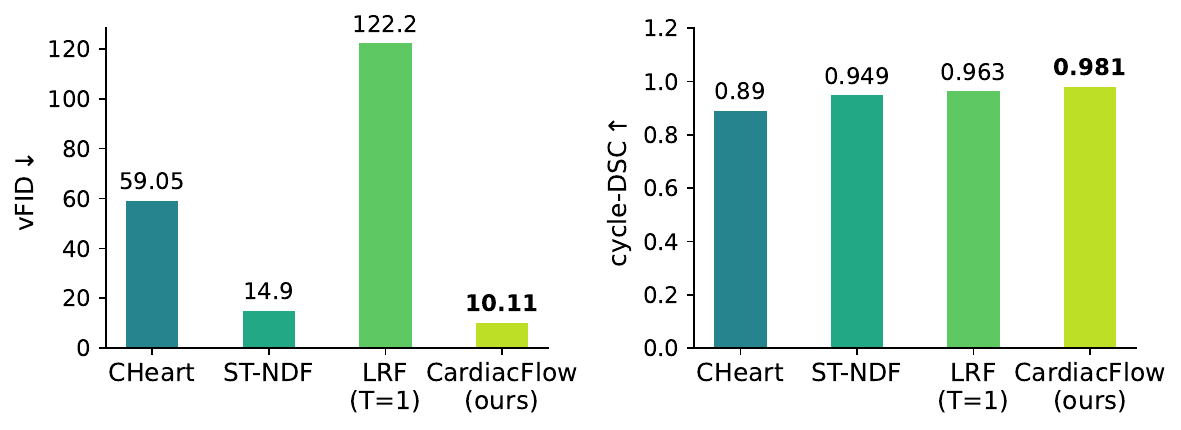}
\caption{3D+t cardiac shape generation results evaluated by vFID and cycle-DSC.}
\label{fig:shape_generation}
\end{minipage}
\end{figure}

We further investigate how the ratio of synthetic data and neural network architectures can affect shape completion. We train LC-U-Net, TransUNet \cite{chen2024transunet}, and SwinUNETR \cite{hatamizadeh2021swinunetr} for 3D cardiac shape completion with synthetic data generated by LRF. Fig. \ref{fig:data_augment} shows that the accuracy of LC-U-Net increases consistently with more synthetic data, until it reaches the best result when 75\% synthetic data are used. LC-U-Net performs better than Transformer-based models \cite{chen2024transunet,hatamizadeh2021swinunetr} since it is data efficient with inductive bias. Regardless of the network architecture, both HD95 and DSC are improved remarkably by LRF-based data augmentation.

\begin{figure}[t]
\centering
\includegraphics[width=0.99\linewidth]{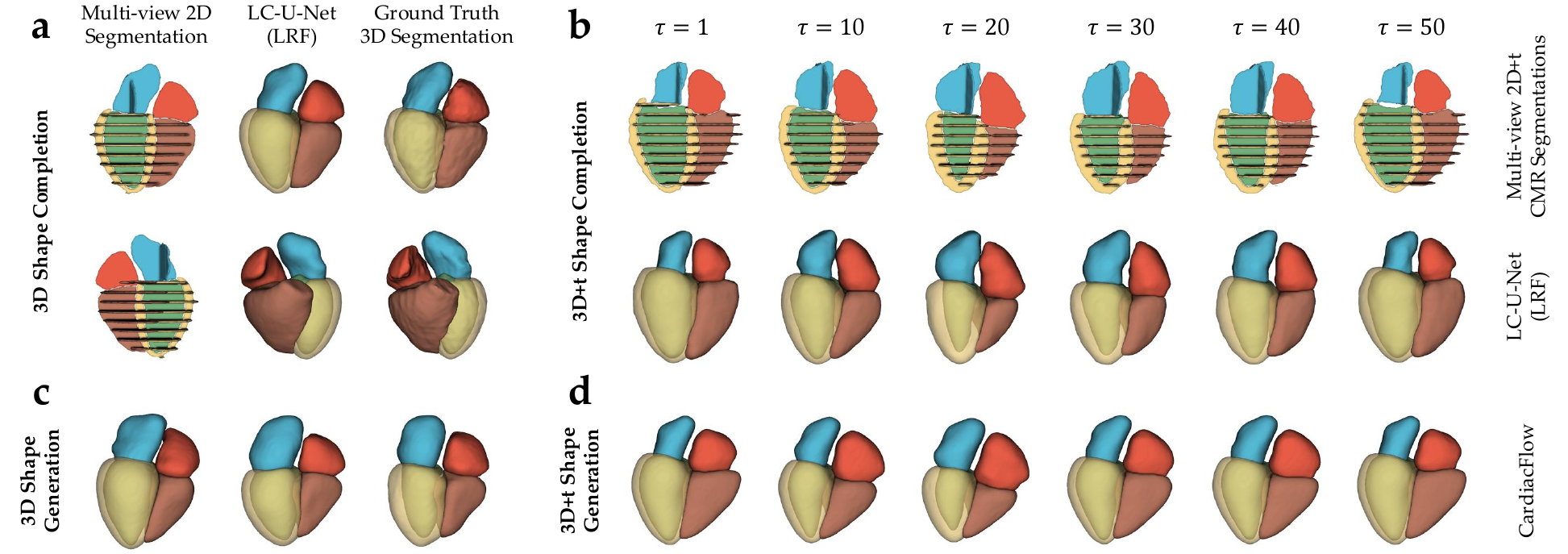}
\caption{Qualitative results for (a) 3D shape completion, (b) 3D+t shape completion, (c) 3D shape generation, and (d) 3D+t shape generation. }
\label{fig:visualization}
\end{figure}

For qualitative results, Fig. \ref{fig:visualization}-a shows a 3D shape completed by LC-U-Net augmented by LRF. Fig. \ref{fig:visualization}-c visualises 3D cardiac shapes generated by LRF. For 3D+t shape completion, we apply LC-U-Net (LRF) to 1,000 UKB data. An example of completed 3D+t cardiac sequence is shown in Fig. \ref{fig:visualization}-b. 
\newline

\begin{figure}[t]
\centering
\includegraphics[width=1.0\linewidth]{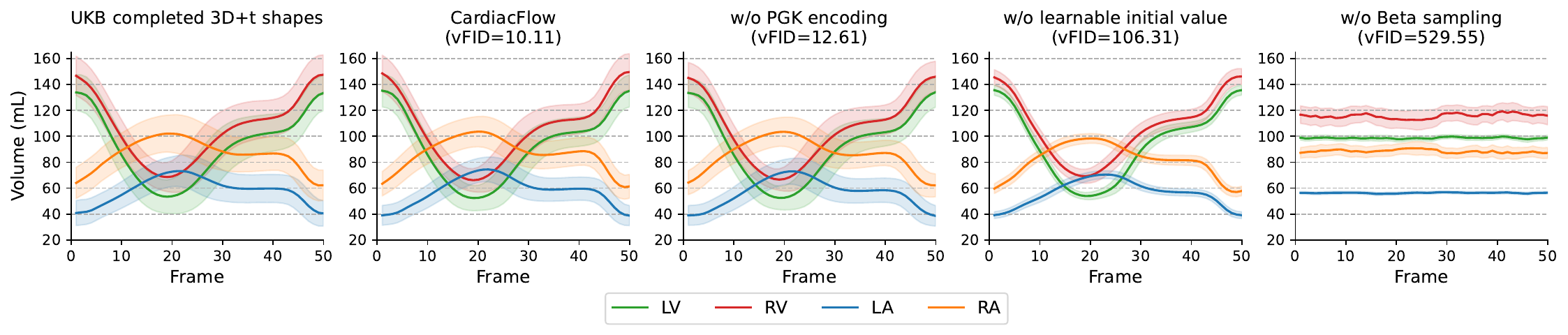}
\caption{Four-chamber volumes of synthetic 3D+t cardiac shapes for ablation studies.}
\label{fig:volume}
\end{figure}

\noindent\textbf{3D+t Shape Generation.} We train CardiacFlow on the UKB dataset with completed 3D+t cardiac shapes for 100 epochs using $\sigma$=1.5 for PGK encoding. CardiacFlow is compared to CHeart \cite{Qiao2023}, spatio-temporal neural distance field (ST-NDF) \cite{sorensen2024stndf}, and a vanilla LRF with one-step generation ($T$=1). For each baseline method, we randomly generate 1,000 samples for evaluation. The quality of generated 3D+t shapes is measured by the volume Fréchet inception distance (vFID) \cite{heusel2017fid}, which is defined as $d_F=\|\mu_1-\mu_2\|^2+\mathrm{tr}(\Sigma_1+\Sigma_2-2(\Sigma_1\Sigma_2)^{\frac{1}{2}})$ between generated 3D+t shapes $\mathcal{N}(\mu_1,\Sigma_1)$ and UKB test set $\mathcal{N}(\mu_2,\Sigma_2)$, where $\mu\in\mathbb{R}^{200}$ is the mean cardiac four-chamber volumes for $M$=50 frames. The periodic consistency is evaluated by cycle-DSC, \emph{i.e.}, the DSC between the start ($\tau$=1) and end ($\tau$=50) frame of a cardiac cycle. The vFID and cycle-DSC are reported in Fig. \ref{fig:shape_generation}, which shows CardiacFlow achieves superior generation quality and periodic consistency than all baselines. CardiacFlow enables high-quality one-step generation, whereas vanilla LRF fails to capture accurate shape distribution in one step. For runtime, CardiacFlow takes 1.611s to generate an entire 3D+t sequence, which is similar to CHeart (1.467s) and ST-NDF (1.472s). An example of 3D+t shapes generated by CardiacFlow is visualised in Fig. \ref{fig:visualization}-d.

We conduct ablation studies on three components of CardiacFlow: (i) The PGK frame encoding $\mathcal{K}_{\sigma}(\tau)$ is substituted by a scalar conditioning variable $\tau$. (ii) The learnable embedding $\epsilon_x$ is replaced by a Gaussian noise so that the initial value is no longer learnable. (iii) During training, the time $t$ is sampled uniformly instead of Beta sampling. The vFID score and cardiac four-chamber volumes averaged over 1,000 generated 3D+t shapes are presented in Fig. \ref{fig:volume}. It indicates that CardiacFlow learns accurate volume distribution from the UKB dataset. The removal of PGK encoding reduces the cycle-DSC from 0.981 to 0.959. Without learnable initial values or beta sampling, CardiacFlow fails to capture the 3D+t shape variation or motion pattern via one-step generation.

\section{Conclusion}
In this work, we introduce flow matching for 3D+t shape completion and generation of four cardiac chambers. A latent rectified flow is employed to generate 3D cardiac shapes for data augmentation, leading to significant improvement on 3D shape completion. A novel CardiacFlow framework is developed to learn efficient 3D+t cardiac shape generation from the UK Biobank dataset. Instead of learning implicit segmentations, in future work, we will explicitly reconstruct 3D+t meshes for cardiac four chambers. In addition, CardiacFlow could be further extended to generate 3D+t cardiac shapes conditioned on demographic and pathological information, as well as model spatio-temporal distribution for various types of medical images and shapes of organs.

\begin{credits}
\subsubsection{\ackname} This work was supported by the EPSRC grants (EP/W01842X/1, EP/Z531297/1) and the BHF New Horizons Grant (NH/F/23/70013). This research was conducted using the UK Biobank Resource under Application Number 18545. We thank all UK Biobank participants and staff.

\subsubsection{\discintname}
The authors have no competing interests to declare that are relevant to the content of this article.
\end{credits}

%
% ---- Bibliography ----
%
% BibTeX users should specify bibliography style 'splncs04'.
% References will then be sorted and formatted in the correct style.
%
% \newpage
\bibliographystyle{splncs04}
\bibliography{ref}

\end{document}